\documentclass[letterpaper, 10 pt, conference]{ieeeconf}

\usepackage[draft]{changes}  
\definechangesauthor[name={Anni}, color=red]{AL}  
\usepackage{booktabs}
\usepackage{tabularx}

\usepackage{bm}
\usepackage{amssymb}
\usepackage{epsfig}
\usepackage{amsmath}
\usepackage{xcolor}
\usepackage{graphicx}
\usepackage{epstopdf}
\usepackage{amsfonts}%
\usepackage{indentfirst}
\usepackage[ruled]{algorithm2e}
\newtheorem{problem}{\textbf{Problem}}
\newtheorem{definition}{\textbf{Definition}}

\newtheorem{theorem}{\rm\textbf{Theorem}}

\newtheorem{remark}{\rm\textbf{Remark}}

\IEEEoverridecommandlockouts
\begin{document}

\title{{\LARGE \textbf{Robust Taylor-Lagrange Control for Safety-Critical Systems with Guaranteed Feasibility 
}}}
\author{Wei Xiao, Christos G. Cassandras and Anni Li
\thanks{W. Xiao and A. Li are with School of Electrical and Electronic Engineering at Nanyang Technological University, Singapore
	\texttt{{\small wei.xiao@ntu.edu.sg, anni.li@ntu.edu.sg}}}
\thanks{C.G. Cassandras is with the Division of Systems Engineering at Boston University, USA
	\texttt{{\small cgc@bu.edu}}}
}

\maketitle

\begin{abstract}


The Control Barrier Function (CBF) method has been widely adopted in solving 
safety-critical control problems.
However, the existence of a CBF is only a sufficient condition for system
safety. The recently proposed Taylor-Lagrange Control (TLC)
method addresses this limitation by providing a necessary and sufficient condition, but it is 
still vulnerable to the
feasibility preservation problem (e.g., due to the inter-sampling effect when time is discretized).
In this paper, we propose a robust TLC (rTLC) method to
address the feasibility preservation problem 
and show how to use it to solve optimal control problems.
Specifically, the
rTLC method expands the safety function to an order higher
than its relative degree using Taylor’s expansion with Lagrange remainder, which allows the control to explicitly
show up at the current time instead of a future time in
the original TLC method. 
This implies that the safety constraint remains satisfied over the entire discretization time interval that follows.
Moreover, the rTLC method naturally addresses the
feasibility preservation problem with only 
a single
hyper-parameter
(the discretization time interval size during implementation),
compared to significantly more parameters for its counterparts. Finally, we illustrate
the effectiveness of the proposed rTLC method through an
adaptive cruise control problem, and compare it with existing
safety-critical control methods.


\end{abstract}

%
\IEEEpeerreviewmaketitle

\section{Introduction}
\label{sec:intro}
Stabilizing dynamical systems while optimizing costs and satisfying (nonlinear) safety constraints has received increasing attention in recent years with 
{the rising use of autonomy, making safety guarantees a challenging issue.
}
Existing methods like optimal control \cite{Bryson1969} \cite{kirk2004optimal} and dynamic programming \cite{bellman1966dynamic} \cite{Denardo2003} are primarily designed for linear systems and constraints, limiting their extensions to nonlinear systems,
{in addition to the fact that they face the ``curse of dimensionality''.
}
Model Predictive Control (MPC) methods \cite{garcia1989model} \cite{Rawlings2018} have been applied to receding horizon safety-critical control, but they are generally computationally heavy, especially for nonlinear systems. Although linearization is possible to improve MPC computational efficiency, such approximations may compromise safety guarantees. Reachability analysis \cite{althoff2014online} \cite{asarin2003reachability} is also widely used to verify system safety, but 
{the computational load it requires becomes prohibitive.
}
In order to address these computational challenges, barrier-based methods have received increasing attention for nonlinear systems.

Barrier functions (BFs) were originally used in optimization problems with inequality constraints \cite{Boyd2004} by 
using
their reciprocal form as part of the cost function. They are also used in machine learning systems to improve safety, such as in safe reinforcement learning \cite{cheng2019end}. However, this method cannot guarantee system safety as the safety constraint is taken as part of the cost (or reward) function, and thus its functionality is similar to a soft constraint.  BFs are also used as Lyapunov-like functions \cite{Wieland2007}, and they have been employed in system verification and control \cite{Tee2009}\cite{Aubin2009}\cite{Prajna2007}\cite{Wisniewski2013}.

Control BFs (CBFs) entail the use of BFs for control (affine) systems. CBFs can trace back to Nagumo's theorem \cite{Nagumo1942berDL} which shows that if the safety function is initially non-negative and its derivative is always non-negative at the safe set boundary, then the safety function remains non-negative for all times. In other words, the safety property can be preserved. For safe control synthesis with CBFs, class $\mathcal{K}$ functions are considered in reciprocal CBFs \cite{Ames2017} and zeroing CBFs \cite{Glotfelter2017} \cite{xiao2021high} to allow the safety function to decrease when the system state is far away from the boundary. {Both CBF forms are fundamentally the same, and they 
{may be}
conservative due to the introduction of class $\mathcal{K}$ functions. In other words, the existence of a CBF is only a \emph{sufficient} condition for system safety. There are different variations of CBFs that have been recently proposed. Stochastic CBFs \cite{clark2021control} have been proposed for stochastic systems, and finite-time convergence CBFs \cite{srinivasan2018control} are used for systems that initially violate the safety constraint. Adaptive CBFs \cite{taylor2020adaptive} \cite{xiao2021adaptive} have been introduced to deal with system uncertainty and noise. To ensure safety for systems and constraints with high relative degree, exponential CBFs (ECBFs) \cite{Nguyen2016} and high-order CBFs (HOCBFs) \cite{xiao2021high} have been proposed. However, the existence of ECBFs or HOCBFs implies the forward invariance of the intersection of a set of safe sets, which is very restrictive and may potentially limit system performance.} The CBF method also introduces additional hyper-parameters, 
{ 
such as those needed to fully characterize 
}
class $\mathcal{K}$ functions
involved in this method,
which may be non-trivial or difficult to tune.

The recently proposed Taylor-Lagrange Control (TLC) method \cite{xiao2025taylor} ensures system safety using Taylor's theorem with Lagrange remainder \cite{taylor1717methodus} \cite{Lagrange1797}. The TLC method has been shown to be a \emph{necessary and sufficient} condition for system safety 
{with considerably fewer hyper-parameters (in fact, only one) than the CBF method. 
}
However, the TLC method is still
vulnerable to the feasibility preservation problem, such as the notorious inter-sampling effect 
{which is a consequence of time discretization to convert a continuous-time system into a sampled-data system.
}
Although we may employ event-triggered methods \cite{taylor2020safety} \cite{xiao2022event} to address the inter-sampling effect, these methods may introduce additional hyper-parameters that are non-trivial to manually tune.

In this paper, we propose a robust Taylor-Lagrange Control (rTLC) method for system safety using Taylor's theorem with Lagrange remainder. The TLC method expands a given safety function to an order higher than the relative degree of this function. This allows the control to show up at the current time instead of 
some
future time which appears in Taylor's expansion. We quantify the bound of the Lagrange remainder in rTLC using a reachable set analysis of the task space set, which eliminates the need to calculate the future time state, control, and the control derivative in an rTLC controller. As a result, the proposed rTLC is shown to naturally address the feasibility preservation problem without additional considerations. Moreover, the rTLC method has only one hyper-parameter in contrast to existing methods, like event-triggered approaches, which require considerably more parameter tuning.

The contributions of this paper are as follows:
\begin{enumerate}
    \item We propose a robust 
    Taylor-Lagrange Control (rTLC) method for system safety, in which robustness is defined in terms of safety preservation with respect to time-discretization errors, and demonstrate its dependence only on the current time instead of future times in Taylor's expansion.

    \item We show that the proposed rTLC method naturally addresses the feasibility preservation problem 
{when time is discretized, i.e., the safety constraint remains satisfied over the
entire discretization time interval that follows.
}
Moreover, this is accomplished with only one hyper-parameter (the discretization time interval size during implementation), thus guaranteeing system safety without additional parameter tuning.

    \item We 
    {illustrate}
    the effectiveness of the proposed rTLC method on an adaptive cruise control problem, and compare it with existing safety-critical control methods.
\end{enumerate}

The remainder of the paper is organized as follows. In Section \ref{sec:prelim}, we provide preliminaries on the TLC method and discuss its connection with CBFs and HOCBFs. We present the rTLC method for nonlinear systems control systems in Section \ref{sec:tlc} with a subsection on how to solve optimal control problems using this method.  A case study and simulation results are shown in Section \ref{sec:case}. We conclude with final remarks and directions for future work in Section \ref{sec:conclusion}.

\section{PRELIMINARIES}
\label{sec:prelim}

In this section, we introduce the background on the Taylor-Lagrange Control (TLC) functions,
and briefly discuss its connection with CBFs. Throughout the paper, we denote $t_0 \geq 0$ as the current time, and $t (> t_0)$ as a future time.

Consider an affine control system 
\begin{equation} \label{eqn:affine}%
\dot {\bm{x}} = f(\bm x) + g(\bm x)\bm u,
\end{equation}
where  $\bm x\in X\in \mathbb{R}^n$ ($X$ is a compact set), $f$ is as defined above, $g:\mathbb{R}^n \rightarrow \mathbb{R}^{n\times q}$ is locally Lipschitz, and $\bm u\in U \subset \mathbb{R}^q$ where $U$ denotes a compact control constraint set:
\begin{equation} \label{eqn:bound}
    U =\{\bm u\in \mathbb{R}^q: \bm u_{min} \leq \bm u\leq \bm u_{max}\},
\end{equation}
where $\bm u_{min} \in \mathbb{R}^q, \bm u_{max} \in \mathbb{R}^q$, and the above constraint is interpreted component-wise. Solutions $\bm x(t)$ of (\ref{eqn:affine}), starting at $\bm x(0)$, $t\geq 0$, are forward complete.

\begin{definition}
    (Class $\mathcal{K}$ function \cite{Khalil2002}) A function $\alpha:[0, a)\rightarrow[0, \infty), a > 0$ is said to belong to class $\mathcal{K}$ if it is strictly increasing and passes the origin.
\end{definition}

\begin{definition}
    (Forward invariance \cite{Ames2017}) A set $C$ is forward invariant for (\ref{eqn:affine}) if its solution $x(t)$ is such that $x(t)\in C, \forall t\geq 0$ for any $x(0)\in C$.
\end{definition}

\begin{definition}
    (Relative degree \cite{Khalil2002}) The relative degree of a function $h(\bm x)$ (or a constraint $h(\bm x)\geq 0$) w.r.t. (\ref{eqn:affine}) is defined as the minimum number of times we need to differentiate $h(\bm x)$ along system (\ref{eqn:affine}) until any control component of $\bm u$ explicitly shows up in the corresponding derivative.
\end{definition}

For a safety constraint $h(\bm x)\geq0$ that has relative
degree $m$ w.r.t system (\ref{eqn:affine}), $h:\mathbb{R}^{n}\rightarrow\mathbb{R}$, we can define a safe set $C$ based on $h$: 
\begin{equation}\label{eqn:C-1}
C := \{\bm x\in \mathbb{R}^n:h(\bm x)\geq 0\},
\end{equation} 
where $h(x) = 0$ at the boundary of $C$ and $h(x) < 0$ outside $C$.
Then, we have the following definition for a Taylor-Lagrange Control (TLC) function:
\begin{definition} [Taylor-Lagrange Control \cite{xiao2025taylor}]\label{def:tlc}
    The function $h:\mathbb{R}^n\rightarrow \mathbb{R}$ is a Taylor-Lagrange Control (TLC) function of relative degree $m$ for system (\ref{eqn:affine}) if
    \begin{equation} \label{eqn:tlc}
       \begin{aligned}
     \sup_{\bm u(\xi)\in U}\left[\sum_{k=0}^{m-1} \frac{L_f^kh(\bm x(t_0))}{k!} (t-t_0)^k + \frac{L_f^mh(\bm x(\xi))}{m!} (t-t_0)^m\right. \\\left.\!+ \frac{L_gL_f^{m-1}h(\bm x(\xi))\bm u(\xi)}{m!} (t\!-\!t_0)^m\right] \!\geq\! 0, \\t_0\in[0,\infty), \xi\in(t_0, t)
\end{aligned}
    \end{equation} for all $\bm x(t_0)\in C$,
    where $L_fh, L_gh$ denote the Lie derivative of $h$ along $f$ and $g$, respectively.
\end{definition}

The TLC exhibits a connection with 
{controllers obtained through the use of 
}
CBFs \cite{Ames2017}. In particular, when $h(\bm x)$ has relative degree $m=1$, then the TLC constraint in (\ref{eqn:tlc}) degenerates to a CBF with a linear class $\mathcal{K}$ function:
\begin{equation}
    L_fh(\bm x(\xi)) + L_gh(\bm x(\xi))\bm u(\xi) + \frac{1}{t-t_0}h(\bm x(t_0)) \geq 0.
\end{equation}
where $\xi = t_0$ in the zero-order hold discretization implementation of such control and $t$ is the next time instant in the discretization. The TLC also extends HOCBFs \cite{xiao2021high} to the complex domain (i.e., the class $\mathcal{K}$ functions in HOCBFs are complex functions) for high relative degree constraints. Moreover, the existence of a TLC is a necessary and sufficient condition for the safety of the system \cite{xiao2025taylor}, while the existence of a CBF is only sufficient.

\begin{theorem}[\cite{xiao2025taylor}]
\label{thm:tlc}
    Given a TLC function $h(\bm x)$ from Def. \ref{def:tlc} with the associated safe set defined in (\ref{eqn:C-1}), if $h(\bm x(t_0))\geq 0$, then any Lipschitz continuous controller $\bm u(\xi)\in K_{tlc}(\bm x(t_0), \bm x(\xi)), \forall t_0\in[0,\infty), \xi\in(t_0,t), t>t_0$ renders the set $C$ forward invariant for system (\ref{eqn:affine}).
\end{theorem}

The TLC can be used for enforcing both system safety and stability \cite{xiao2025taylor}. We may first discretize time, and then employ a zero-order hold implementation within each time interval. We take $t_0$ as the current time instant, and define $t=t_1$ in Def. \ref{def:tlc}, where $t_1$ is the next time instant of implementing the control update.
This allows us to transform the original nonlinear constrained control problem into a sequence of quadratic programs (QPs). The inter-sampling effect of the TLC 
(i.e, the effect of holding the control constant over a time step)
can be addressed by event-triggered methods \cite{xiao2025taylor}, 
{i.e., by triggering the next QP solution to occur at an appropriately defined event rather than a predefined discretization time step.
}
However, event-triggered methods may have too many hyper-parameters that are hard to tune. In this paper, we address the inter-sampling effect using the proposed robust TLC method that 
{involves a single hyper-parameter. 
}

\section{Robust Taylor-Lagrange Control}
\label{sec:tlc}

In this section, we propose the robust Taylor-Lagrange Control (rTLC) for obtaining state-feedback safe control policies for nonlinear control systems while addressing the inter-sampling effect.
{
By ``robust'' we mean that the control is guaranteed to remain feasible and ensure system safety despite the fact that it is held constant during a time interval, which in general ceases to guarantee the forward invariance property in Theorem \ref{thm:tlc}.} We also discuss how to solve constrained optimal control problems using the proposed rTLC. 
There may be other reasons for which feasibility preservation can fail (e.g., due to control limits), but we limit ourselves here to the inter-sampling effect only.

\subsection{Robust Taylor-Lagrange Control}
Suppose the relative degree of the safety constraint $h(\bm x)\geq 0$ is $m$ with respect to (\ref{eqn:affine}), and the function $h(\bm x)$ is $m+1$ times differentiable. {\it We first treat $\bm u(t)$ as a continuously differentiable control signal, and then find a constant $\bm u$ over the current time interval $[t_0, t]$ that satisfies the corresponding rTLC constraint.}

Given a TLC as in Def. \ref{def:tlc}, we notice that the control explicitly shows up in the term $\frac{L_gL_f^{m-1}h(\bm x(\xi))\bm u(\xi)}{m!} (t\!-\!t_0)^m$, where $\xi\in (t_0, t)$ is hard to determine. Note that $\bm u(\xi)$ is the control at time $\xi$ instead of the current time $t_0$, which makes it hard to find a control policy for the current time $t_0 < \xi$. Although we can employ a zero-order implementation of the control
such that $t = t_1$ ($t_1$ denotes the next implementation time instant) and $\bm u(t_0) = \bm u(\xi)$, this will generally introduce some errors. We may further employ event-triggered methods \cite{xiao2025taylor} to address the inter-sampling effect due to the zero-order hold error; however, there are usually too many hyper-parameters that are hard to tune. 
{In order to avoid this issue, we propose to expand the function $h(\bm x)$ to an order of $m+1$ instead of $m$ as described next.
}

\textbf{Higher-order TLC.} Let $\Delta t = t-t_0$. We expand the function $h(\bm x)$ to an order of $m+1$ instead of $m$ in Taylor's theorem 
{so that (\ref{def:tlc}) becomes:
}
\begin{equation} \label{eqn:rtlc1}
       \begin{aligned}
     \sum_{k=0}^{m-1} \frac{L_f^kh(\bm x(t_0))}{k!} {\Delta t}^k + \frac{L_f^mh(\bm x(t_0))}{m!} {\Delta t}^m \\\!+ \frac{L_gL_f^{m-1}h(\bm x(t_0))\bm u(t_0)}{m!} {\Delta t}^m \\+ R(\bm x(\xi), \bm u(\xi), \dot{\bm u}(\xi), \Delta t)\!\geq\! 0, t_0\in[0,\infty), \xi\in(t_0, t),
\end{aligned}
\end{equation}
where the Lagrange remainder is defined as
\begin{equation} \label{eqn:tlc-remain}
\begin{aligned}
    R(\bm x(\xi), \bm u(\xi), \dot{\bm u}(\xi), \Delta t) = \frac{L_f^{m+1}h(\bm x(\xi)) }{(m+1)!} {\Delta t}^{m+1} \\+ \frac{ L_gL_f^{m}h(\bm x(\xi))\bm u(\xi)+L_fL_gL_f^{m-1}h(\bm x(\xi))\bm u(\xi)}{(m+1)!} {\Delta t}^{m+1}  \\+ \frac{L_g^2L_f^{m-1}h(\bm x(\xi))\bm u^2(\xi) + L_gL_f^{m-1}h(\bm x(\xi))\dot{\bm u}(\xi)}{{(m+1)}!} {\Delta t}^{m+1} 
\end{aligned}
\end{equation}
Note that $\dot{\bm u}{(\xi)}$ shows up in the above remainder because the relative degree of $h(\bm x)$ is $m$ (the first time when the control ${\bm u}{(\xi)}$ shows up when taking derivatives of $h(\bm x)$), 
while we expand $h(\bm x)$ to a $(m+1)^{th}$ order. Since $\bm u$ is treated as a continuously differentiable control, it holds that $\dot{\bm u} \ne 0$.


In (\ref{eqn:tlc-remain}), the function $h(\bm x)$ is $m+1$ times differentiable, and $\bm x\in X, \bm u\in U$, where $X, U$ are compact sets. In order to ensure that $R(\bm x(\xi), \bm u(\xi), \dot{\bm u}(\xi), \Delta t)$ is bounded, we need to check whether $\dot{\bm u}(\xi)$ is bounded (as $\dot{\bm u}\ne 0$).

\textbf{Boundedness of $\dot{\bm u}(\xi)$.} The control set $U$ is defined as in (\ref{eqn:bound}). We need to ensure that $\bm u\in U$ when $\dot{\bm u}$ is introduced. In other words, $\dot{\bm u}$ cannot be unbounded. Let $\bm u = (u_1.\dots, u_q), \bm u_{min} = (u_{1,min},\dots, u_{q,min}), \bm u_{max} = (u_{1,max},\dots, u_{q,max})$, where $u_{i,min} \in\mathbb{R}, u_{i,max} \in\mathbb{R}, \forall i\in\{1,\dots, q\}$. In order to ensure $\bm u\in U$ with $\dot{\bm u}$ governing the evolution of control $\bm u$, 
we introduce two first order 
TLC functions
for each $i\in\{1,\dots, q\}$:
\begin{equation}
\begin{aligned}
    h_{i,1}(\bm u(\xi)) = u_{i,max} - u_{i},\\
    h_{i,2}(\bm u(\xi)) = u_{i} - u_{i,min},
\end{aligned}
\end{equation}

In order to ensure the two functions above become TLCs, the following constraints have to be satisfied:
\begin{equation}
\begin{aligned}
    L_fh_{i,1}(u_i(\xi)) + L_gh_{i,1}(u_i(\xi))\dot{u}_i(\xi) + \frac{1}{\Delta t} {h}_{i,1}(u_i(\xi)) \geq 0,\\
    L_fh_{i,2}(u_i(\xi)) + L_gh_{i,2}(u_i(\xi))\dot{u}_i(\xi) + \frac{1}{\Delta t} {h}_{i,2}(u_i(\xi))\geq 0,
\end{aligned}
\end{equation}

Combining the last two equations, it follows that $\dot{u}_i(\xi)$ is bounded by:
\begin{equation}
\begin{aligned}
    -\frac{1}{\Delta t}(u_i(\xi) - u_{i,min})\leq \dot u_i(\xi) \leq \frac{1}{\Delta t}(u_{i,max} - u_{i}(\xi)).
\end{aligned}
\end{equation}
The above equation can be rewritten in vector form:
\begin{equation} \label{eqn:du_bound}
\begin{aligned}
    -\frac{1}{\Delta t}(\bm u(\xi) - \bm u_{min})\leq \dot {\bm u}(\xi) \leq \frac{1}{\Delta t}(\bm u_{max} - \bm u(\xi)),
\end{aligned}
\end{equation}
where the inequality above is interpreted component-wise.

\textbf{Boundedness of $R(\bm x(\xi), \bm u(\xi), \dot{\bm u}(\xi), \Delta t)$.} Since $\dot {\bm u}(\xi)$ has to be bounded as shown in (\ref{eqn:du_bound}) in order to ensure $\bm u\in U$, 
it follows that $R(\bm x(\xi), \bm u(\xi), \dot{\bm u}(\xi), \Delta t)$ is also bounded as $\bm x\in X, \bm u\in U$, where $X, U$ are compact sets.

Let $S_R(\bm x(\xi), \bm u(\xi), \dot{\bm u}(\xi))$ denote the reachable set of $R(\bm x(\xi), \bm u(\xi), \dot{\bm u}(\xi))$ within the time interval $(t_0, t)$, where $\xi \in (t_0, t)$. Then we can find the minimum value $R_{min}(\Delta t)$ of $R(\bm x(\xi), \bm u(\xi), \dot{\bm u}(\xi), \Delta t)$ by:
\begin{equation} \label{eqn:lb}
    R_{min}(\Delta t) \!=\! \min_{(\bm x(\xi), \bm u(\xi), \dot{\bm u}(\xi))\in S_R(\bm x(\xi), \bm u(\xi), \dot{\bm u}(\xi))} \!\!\!\!\!\!\!\!\hspace{-8mm}R(\bm x(\xi), \bm u(\xi), \dot{\bm u}(\xi),\Delta t).
\end{equation}

Since $h(\bm x)$ is $m+1$ times differentiable, $\bm u\in U, \bm x\in X$, where $U, X$ are compact sets, and $\dot{\bm u}$ is also bounded as in (\ref{eqn:du_bound}), it follows that $R_{min}$ must exist. Let $U_{\dot u}$ be defined by:
\begin{equation}
\begin{aligned}
    U_{\dot u} := \{\dot{\bm u}(\xi)\in\mathbb{R}^q:-\frac{1}{\Delta t}(\bm u(\xi) - \bm u_{min})\leq \dot {\bm u}(\xi) \leq \\\frac{1}{\Delta t}(\bm u_{max} - \bm u(\xi)).\}
\end{aligned}
\end{equation}
Since $\bm u(\xi)\in U$, we have that $U_{\dot u}$ is a subset of 
\begin{equation}
\begin{aligned}
    U_{\dot u, sup} := \{\dot{\bm u}(\xi)\in\mathbb{R}^q:-\frac{1}{\Delta t}(\bm u_{max} - \bm u_{min})\leq \dot {\bm u}(\xi) \leq \\\frac{1}{\Delta t}(\bm u_{max} - \bm u_{min}.\}
\end{aligned}
\end{equation}

Finally, we have that $R_{min}$ in (\ref{eqn:lb}) is lower bounded by:
\begin{equation}
    R_{min,l}(\Delta t) \geq \min_{(\bm x(\xi), \bm u(\xi), \dot{\bm u}(\xi))\in (X, U, U_{\dot u, sup})} \hspace{-9mm}R(\bm x(\xi), \bm u(\xi), \dot{\bm u}(\xi),\Delta t).
\end{equation}
The right-hand side of the above equation can  always be obtained as we do not explicitly need to calculate the reachable set $S_R(\bm x(\xi), \bm u(\xi), \dot{\bm u}(\xi))$. Deriving a tight reachable set for $S_R(\bm x(\xi), \bm u(\xi), \dot{\bm u}(\xi))$ could significantly reduce the conservativeness of our proposed robust TLC method, and thus is worth a future study.

Now, we are ready to define the robust TLC as follows:
\begin{definition} [Robust Taylor-Lagrange Control]\label{def:rtlc}
    Let $R_{min}(\Delta t)$ be defined as in (\ref{eqn:lb}). The function $h:\mathbb{R}^n\rightarrow \mathbb{R}$ is a robust Taylor-Lagrange Control (rTLC) function of relative degree $m$ for system (\ref{eqn:affine}) if
    \begin{equation} \label{eqn:rtlc}
       \begin{aligned}
     \sup_{\bm u(t_0)\in U} \min_{\tau\in(0,\Delta t]}\left[\sum_{k=0}^{m-1} \frac{L_f^kh(\bm x(t_0))}{k!} {\tau}^k + \frac{L_f^mh(\bm x(t_0))}{m!} {\tau}^m\right. \\\left.\!+ \frac{L_gL_f^{m-1}h(\bm x(t_0))\bm u(t_0)}{m!} {\tau}^m + R_{min}(\tau)\right] \!\geq\! 0, 
\end{aligned}
    \end{equation} for $t_0\in[0,\infty),$ and for all $\bm x(t_0)\in C$.
\end{definition}
Note that in Def. \ref{def:rtlc}, we exclude the case when $\tau = 0$ as the constraint in (\ref{eqn:rtlc}) is exactly the original constraint $h(\bm x(t_0))\geq 0$ when $\tau = 0$ and no control input will show up.

{It is important to note that}
the difference between the rTLC in Def. \ref{def:rtlc} and the TLC in Def. \ref{def:tlc} lies in that there is no $\xi \in (t_0, t)$ present in Def. \ref{def:rtlc}. Therefore, the constraint is now defined over $\bm u(t_0)$ instead of $\bm u(\xi)$. Moreover, (\ref{eqn:rtlc}) only depends on $\bm x(t_0)$ without any dependence on $\bm x(\xi)$. This property significantly facilitates the implementation of the rTLC method.

Given a rTLC as in Def. \ref{def:rtlc}, we can define a state feedback controller that satisfies (\ref{eqn:rtlc}):
\begin{equation}
\begin{aligned}
    K_{rtlc}(\bm x(t_0)) = \{\bm u(t_0)\in U:  \min_{\tau\in[0,\Delta t]}\sum_{k=0}^{m-1} \frac{L_f^kh(\bm x(t_0))}{k!} {\tau}^k  \\+ R_{min}(\tau) \\+ \frac{L_f^mh(\bm x(t_0))}{m!} {\tau}^m + \frac{L_gL_f^{m-1}h(\bm x(t_0))\bm u(t_0)}{m!} {\tau}^m \geq 0
    \}.
\end{aligned}
\end{equation}

\begin{theorem}
\label{thm:rtlc}
    Given a rTLC function $h(\bm x)$ from Def. \ref{def:rtlc} with the associated safe set defined as in (\ref{eqn:C-1}), if $h(\bm x(t_0))\geq 0$, then any Lipschitz continuous controller $\bm u(t_0)\in K_{rtlc}(\bm x(t_0))$ ensures $h(\bm x(\tau))\geq 0$ $\forall \tau\in[t_0, t]$. Moreover, if $\bm u(t_0)\in K_{rtlc}(\bm x(t_0))$ for all $t_0\in [0,\infty)$, then the set C  is forward invariant for system (\ref{eqn:affine}).
\end{theorem}
\noindent\textit{Proof:} By (\ref{eqn:lb}), we have that \begin{equation} 
       \begin{aligned}
     \min_{\tau\in[0,\Delta t]}\sum_{k=0}^{m-1} \frac{L_f^kh(\bm x(t_0))}{k!} {\tau}^k + \frac{L_f^mh(\bm x(t_0))}{m!} {\tau}^m \\\!+ \frac{L_gL_f^{m-1}h(\bm x(t_0))\bm u(t_0)}{m!} {\tau}^m + R(\bm x(\xi), \bm u(\xi), \dot{\bm u}(\xi),\tau)\!\geq\! \\\min_{\tau\in[0,\Delta t]}\sum_{k=0}^{m-1} \frac{L_f^kh(\bm x(t_0))}{k!} {\tau}^k   + \frac{L_f^mh(\bm x(t_0))}{m!} {\tau}^m + R_{min}(\tau) \\+ \frac{L_gL_f^{m-1}h(\bm x(t_0))\bm u(t_0)}{m!} {\tau}^m, t_0\in[0,\infty), \xi\in(t_0, t_0+\tau),
\end{aligned}
\end{equation}
Since  $\bm u(t_0)\in K_{rtlc}(\bm x(t_0))$, we have that \begin{equation} 
       \begin{aligned}
     \sum_{k=0}^{m-1} \frac{L_f^kh(\bm x(t_0))}{k!} {\tau}^k + \frac{L_f^mh(\bm x(t_0))}{m!} {\tau}^m \\\!+ \frac{L_gL_f^{m-1}h(\bm x(t_0))\bm u(t_0)}{m!} {\tau}^m + R(\bm x(\xi), \bm u(\xi), \dot{\bm u}(\xi),\tau)\!\geq\! 0, \\ t_0\in[0,\infty), \tau\in[0,\Delta t],\xi\in(t_0, t_0+\tau),
\end{aligned}
\end{equation}

As the left side of the above constraint is the Taylor's expansion of $h(\bm x(\tau)), \tau\in[t_0,t]$ (as $\Delta t = t-t_0$) with Lagrange remainder, we have
\begin{equation}
    h(\bm x(\tau)) \geq 0, ~~\tau\in[t_0, t].
\end{equation}
 When we move $t_0$ within $[0,\infty)$, we can ensure that $h(\bm x(t))\geq 0, \forall t\in [t_0, \infty)$. Thus, the set C  is forward invariant for system (\ref{eqn:affine}). $\hfill \blacksquare$

 Note that we have employed a constant $\bm u(t_0)$ to substitute the continuous $\bm u(\tau), \tau\in[t_0, t]$ in the rTLC. This is 
 consistent with
 how we implement the control in real scenarios, i.e., with piece-wise constant controls.

\begin{remark}[rTLC addresses the inter-sampling effect] As shown in the proof of Thm. \ref{thm:rtlc}, any controller $\bm u(t_0)$ that satisfies the rTLC constraint in (\ref{eqn:rtlc}) ensures $h(\bm x(\tau))\geq 0$ for all $\tau\in [t_0, t]$. Therefore, the controller $\bm u(t_0)$ obtained at the current state $\bm x(t_0)$ can naturally address the inter-sampling effect. Moreover, in (\ref{eqn:rtlc}), the only parameter we need to consider is $\Delta t$, whereas other approaches (like event-triggered methods \cite{taylor2020safety} \cite{xiao2022event}) that address the inter-sampling effect require tuning a significantly larger number of parameters.
\end{remark}

\subsection{Optimal Control with rTLC}
\label{sec:oc}

Consider a constrained optimal control problem for system (\ref{eqn:affine}) with a cost defined as:
\begin{equation}\label{eqn:cost}
J(\bm u(t)) = \int_{0}^{T}\mathcal{C}(||\bm u(t)||)dt
\end{equation}
where $||\cdot||$ denotes the 2-norm of a vector. $T$ denotes the final time, and  $\mathcal{C}(\cdot)$ is a strictly increasing function of its argument (such as the energy consumption function $\mathcal{C}(||\bm u(t)||) = ||\bm u(t)||^2$).

We also wish the state $\bm x$ of system (\ref{eqn:affine}) to reach a desired state $\bm x_d\in\mathbb{R}^n$ at the final time $T$. Formally, we wish to achieve the following:
\begin{equation} \label{eqn:convergence}
    \min_{\bm u} ||\bm x(T) - \bm x_d||.
\end{equation}

Assume a safety constraint 
\begin{equation} \label{eqn:safety}
h(\bm x) \geq 0,
\end{equation}
with relative degree $m$ that has to be satisfied by system (\ref{eqn:affine}). 

Thus, we need to consider the following constrained optimal control problems:
\begin{problem} \label{problem}
Our objective is to find a controller $\bm u^*$ for (\ref{eqn:affine}) by solving the following optimization problem:
\begin{equation}
\begin{aligned}
    \bm u^* &= \arg\min_{\bm u\in U} \int_{0}^{T}\mathcal{C}(||\bm u(t)||)dt + w||\bm x(T) - \bm x_d||,\\
    &\text{ s.t. (\ref{eqn:safety}),} 
\end{aligned}
\end{equation}
where $w > 0$. We may of course consider multiple safety constraints of the form (\ref{eqn:safety}) in the above. 
\end{problem}

\textbf{Approach:} We use a rTLC to enforce the satisfaction of the safety constraint (\ref{eqn:safety}) for all times. We may also employ a Taylor-Lagrange Stability (TLS) function \cite{xiao2025taylor} defined below to enforce the state convergence objective (\ref{eqn:convergence}).
\begin{definition}  [Taylor-Lagrange Stability \cite{xiao2025taylor}]\label{def:tls}
	 A continuously differentiable function $V: \mathbb{R}^n\rightarrow \mathbb{R}$ is a Taylor-Lagrange Stability (TLS) function of relative degree $m$ for system (\ref{eqn:affine}) if there exist constants $c_1 >0, c_2>0$  such that, for $\forall \bm x\in \mathbb{R}^n$,
	\begin{equation}\label{eqn:tls}
    \begin{aligned}
    &c_1||\bm x||^2 \leq V(\bm x) \leq c_2 ||\bm x||^2,\\
	&\inf_{\bm u(\xi)\in U}\left[\sum_{k=0}^{m-1} \frac{L_f^kV(\bm x(t_0))}{k!} \Delta t^k + \frac{L_f^mV(\bm x(\xi))}{m!} \Delta t^m\right. \\&\left.\!+ \frac{L_gL_f^{m-1}V(\bm x(\xi))\bm u(\xi)}{m!} \Delta t^m\right] \!\leq\! 0, t_0\in[0,\infty), \xi\in(t_0, t).
    \end{aligned}
	\end{equation}
\end{definition}
Any control that satisfies the above TLS constraint can ensure the system state converging to the desired state. We may also introduce a similar method as rTLC to address the inter-sampling effect. However, since we usually use a TLS to enforce a soft constraint when it conflicts with rTLC, we omit this inter-sampling error in practice (i.e., setting $\xi = t_0$ in Def. \ref{def:tls}). 
We directly take the cost (\ref{eqn:cost}) as the objective to reformulate Problem \ref{problem} as a sequence of Quadratic Programs (QPs) where the control $\bm u$ is the decision variable. Since the rTLC only needs to consider the current state $\bm x(t_0)$ without $\bm x(\xi), \xi \in (t_0, t)$, we can directly solve Problem \ref{problem} by solving the QPs.

\section{CASE STUDY AND RESULTS}
\label{sec:case}
In this section, we present a case study 
{for optimal and safe control of vehicles engaged in Adaptive Cruise Control (ACC).
}
All the computations were conducted in MATLAB, and we used quadprog to solve QPs and ode45 to integrate the dynamics.

Consider the ACC problem with the vehicle dynamics having the form:
\begin{equation} \label{eqn:simpledynamics}
\left[\begin{array}{c} 
\dot v(t)\\
\dot z(t)
\end{array} \right]=
\left[\begin{array}{c}  
\frac{-F_r(v(t))}{M}\\
v_0 - v(t)
\end{array} \right] + 
\left[\begin{array}{c}  
\frac{1}{M}\\
0
\end{array} \right]u(t),
\end{equation}
where $v(t)$ denotes the velocity of the ego vehicle along its lane, $z(t)$ denotes the along-lane distance between the ego and its preceding vehicle, $v_0 > 0$ denotes the speed of the preceding vehicle, $M$ is the mass of the ego vehicle, and $u(t)$ is the control input of the controlled vehicle. $F_{r}(v(t))$ denotes the resistance force, which is expressed \cite{Khalil2002} as:
\begin{equation}\label{eqn:resistence}
F_{r}(v(t)) = f_0sgn(v(t)) + f_1v(t) + f_2 v^2(t),
\end{equation}
where $f_0 > 0, f_1 > 0$ and $f_2 > 0$ are scalars determined empirically. The first term in $F_{r}(v(t))$ denotes the Coulomb friction force, the second term denotes the viscous friction force and the last term denotes the aerodynamic drag.

\textbf{Objective 1.} We wish to minimize the acceleration of the ego vehicle through the quadratic form:
\begin{equation}
    \min_u \int_0^T \left(\frac{u-F_r(v)}{M}\right)^2 dt.
\end{equation}

\textbf{Objective 2.} The ego vehicle seeks to achieve a desired speed $v_d > 0$.

\textbf{Safety.} We require that the distance $z(t)$ between the controlled vehicle and its immediately preceding vehicle be greater than a constant $\delta > 0$ for all the times, i.e.,
\begin{equation} \label{eqn:safety_acc}
z(t) \geq c,\quad\forall t\geq 0,
\end{equation}
where $c > 0$.

\textbf{Control limitations.} We also consider a control constraint $u_{min}\leq u(t)\leq u_{max}$, where $u_{min}=-c_dMg, ~u_{max} = c_aMg, ~g = 9.81m/s^2, ~c_a > 0, ~c_d > 0$ for (\ref{eqn:simpledynamics}). 

The ACC problem is to find a control policy that achieves Objectives 1 and 2 subject to the safety constraint and control bounds.
The relative degree of (\ref{eqn:safety_acc}) is two, and we use 
four separate control methods: a second order HOCBF, TLC,  event-triggered TLC, and the proposed rTLC. We define the safety constraint $h(\bm x) = z - c \geq 0$ and obtain the HOCBF and TLC constraint inequalities as follows:
\begin{equation} \label{eqn:hocbf-case}
\begin{aligned}
    L_f^2h(\bm x) + L_gL_fh(\bm x)u + p_1q_1L_fh(\bm x) h^{q_1-1}(\bm x) \\+ p_2(L_fh(\bm x) + p_1h^{q_1}(\bm x))\geq 0. 
\end{aligned}
\quad\text{HOCBF}
\end{equation}
where $p_1 > 0, p_2 > 0, q_1 \geq 1, q_2\geq 1$ are parameters to be designed.

\begin{equation} \label{eqn:tlc-case}\begin{aligned}
      L_f^2h(\bm x) + L_gL_fh(\bm x)u +  \frac{2}{{\Delta t}}L_fh(\bm x) \\+ \frac{2}{{\Delta t}^2}h(\bm x) \geq 0.
\end{aligned}
\qquad\text{TLC}
\end{equation}
{Next, for the event-triggered TLC, we need to define state bounds so that the events triggering a new QP solution occur when these bounds are reached:
} 
$$S(\bm x) = \{\bm y\in \mathbb{R}^2: \bm x - \bm x_{lower}\leq \bm y\leq  \bm x + \bm x_{up}\},$$ 
where $\bm x_{lower} \in\mathbb{R}^2, \bm x_{up} \in\mathbb{R}^2$.

{Finally, the rTLC control must satisfy:}
\begin{equation} \label{eqn:rtlc-case}\begin{aligned}
      L_f^2h(\bm x(t_0)) + L_gL_fh(\bm x)u(t_0) + \min_{\tau\in(0,\Delta t]} \left[\right.\frac{2}{{\tau}}L_fh(\bm x(t_0)) \\+ \frac{2}{{\tau}^2}h(\bm x(t_0)) + R(\bm x(\xi), u(\xi), \dot u(\xi),\tau) \left.\right]\geq 0.
\end{aligned}
\end{equation}
where $\xi \in (t_0, t_0+\tau)$ and 
\begin{equation}
\begin{aligned}
   R(\bm x(\xi), u(\xi), \dot u(\xi),\tau) \!=\! \frac{\tau}{3} ((L_f^3h(\bm x(\xi)) \!+\! L_gL_f^2h(\bm x(\xi))) u(\xi) \\+ L_fL_gL_fh(\bm x(\xi))u(\xi) + L_g^2L_fh(\bm x(\xi))u^2(\xi) \\+ L_gL_fh(\bm x(\xi)) \dot u(\xi)), \xi \in (t_0, t_0+\tau).
\end{aligned}
\end{equation}
where $\tau\in(0,\Delta t]$.

Following (\ref{eqn:du_bound}), the above equation can be rewritten as
\begin{equation}
\begin{aligned}
   R(\bm x(\xi), &u(\xi), \dot u(\xi),\tau) \!=\! \frac{\tau}{3} (-\frac{\dot u(\xi) \!-\! f_1 u(\xi) -2f_2v(\xi)u(\xi)}{M}) \\ &\geq \frac{\tau}{3} (\frac{ -\frac{1}{\tau}(u_{max} - u(\xi)) + f_1 u(\xi) +2f_2v(\xi)u(\xi)}{M})\\
   &\geq \frac{\tau}{3} (\frac{ -\frac{1}{\tau}(u_{max} - u_{min}) + f_1 u_{min} +2f_2v(\xi)u_{min}}{M}) \\
   &> \frac{\tau}{3} (\frac{ -\frac{1}{\tau}(u_{max} - u_{min}) + f_1 u_{min}}{M}) \\&\qquad + \frac{\tau}{3} (\frac{  2f_2(v(t_0) + \Delta t u_{max}/M)u_{min}}{M}) \\
   &\geq \frac{\Delta t}{3} (\frac{ -\frac{1}{\Delta t}(u_{max} - u_{min}) + f_1 u_{min}}{M}) \\&\qquad + \frac{\Delta t}{3} (\frac{  2f_2(v(t_0) + \Delta t u_{max}/M)u_{min}}{M}),
\end{aligned}
\end{equation}
where $v_{max}$ is the maximum speed. We then define $R_{min}$ to be the lower bound in the above equation. Note that we can find a closed-form solution for the $\min$ part (excluding $R(\bm x(\xi), u(\xi), \dot u(\xi),\tau)$) in (\ref{eqn:rtlc-case}) as it is a quadratic function of $\frac{1}{\tau}$.
{We also employ a TLS with relative degree one to enforce the desired speed as in (\ref{eqn:tls})}.

All the simulation parameters are shown in Table \ref{table:param}, in which $dt$ is the time step used when discretizing the system dynamics (\ref{eqn:simpledynamics}) 
{
(not involved with the implementation of rTLC),
whereas $\Delta t$ is the time interval dictating control updates 
(the control held constant over this interval before it is updated through the next QP solution).
}
As long as $dt \leq \Delta t$, we can easily show that the safety of the system can be guaranteed. 

\begin{table}
 	\caption{Simulation parameters for ACC}\label{table:param}
 	\begin{center}
 		\begin{tabular}{|c||c||c|c||c||c|}
 			\hline
Parameter & Value & Units &Parameter & Value & Units\\
\hline
\hline
$v(0)$ & 24& $m/s$&	$z(0)$ & 90& $m$\\
\hline
$v_{0}$ & 13.89& $m/s$ & $v_d$ & 24& $m/s$\\
\hline
$M$ & 1650& $kg$ &g & 9.81& $m/s^2$\\
\hline
$f_0$ & 0.1& $N$ &$f_1$ & 5& $Ns/m$\\
\hline
$f_2$ & 0.25& $Ns^2/m$ &$c$ & 10& $m$\\
\hline
$\Delta t$ & 0.1& $s$&	$dt$ & 0.1& $s$\\
\hline
$c_a$ & 0.4& unitless&$c_d$ & 0.7& unitless\\
\hline
$\bm x_{lower}$ & (0.5, 1)& unitless&$\bm x_{up}$ & (0.5,1)& unitless\\
\hline
 		\end{tabular}
 	\end{center}
 	
 \end{table}

\begin{figure}[thpb]
	\centering
	\includegraphics[scale=0.4]{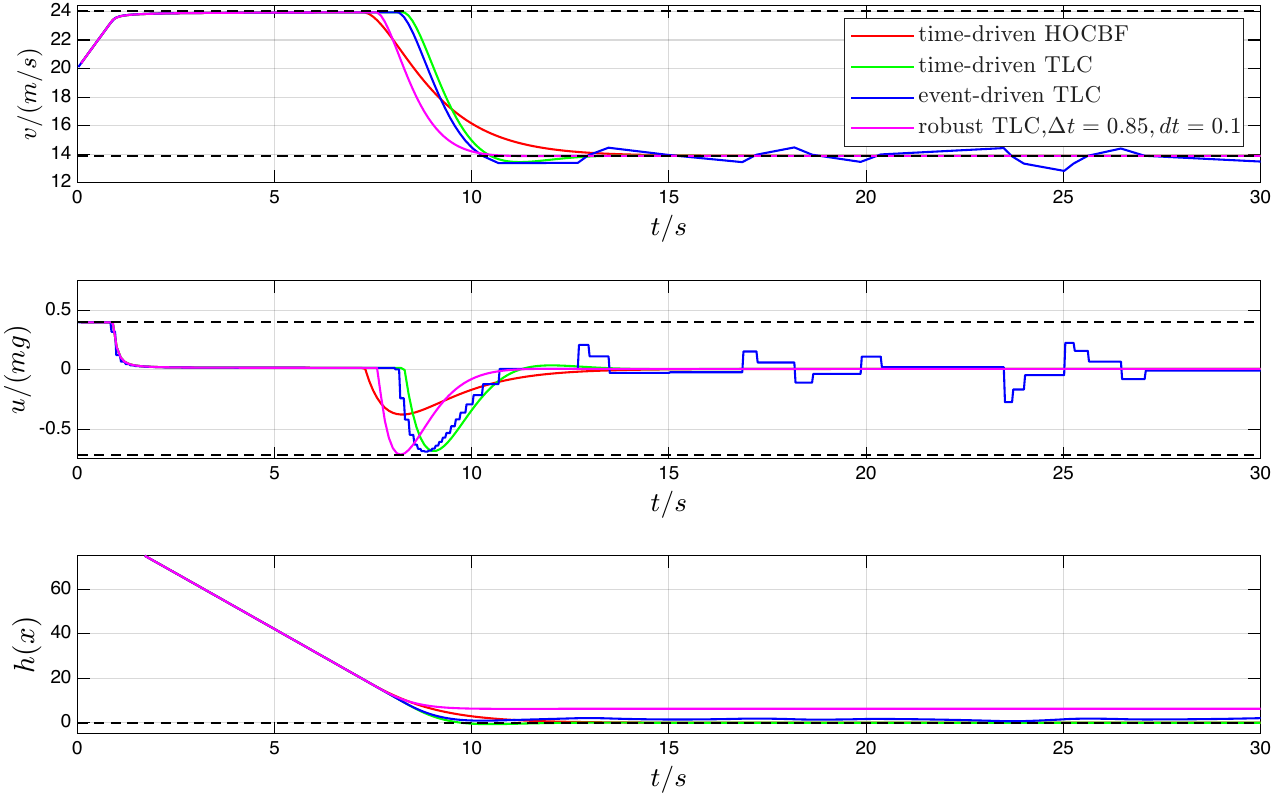}
	\caption{Comparison of speed, control and safety function $h(\bm x)$ profiles between between time-driven HOCBF, time-driven TLC, event-driven TLC, and the proposed rTLC ($\Delta t = 0.85, dt = 0.1$).}
	\label{fig:time-compare}	
\end{figure}

 Simulation results are shown in Fig \ref{fig:time-compare}. The HOCBF, event-driven TLC and rTLC can all guarantee the vehicle safety, while the time-driven TLC may violate the safety constraint due to the inter-sampling effect
 (see Table \ref{tab:cmp}).
 However, the rTLC may be conservative as the vehicle stays unnecessarily far away from the preceding vehicle, as shown with a larger $h(\bm x)$ steady value for rTLC in the last plot of Fig. \ref{fig:time-compare}. We found that decreasing $\Delta t$ can significantly reduce this conservativeness, as shown in Table \ref{tab:cmp}. In terms of computational cost, note that the event-driven TLC is a little more computationally expensive than the other methods. 
 
 {
 \textbf{Number of parameters to tune.}
Observe that the HOCBF-based controller heavily depends on the parameters in the class $\mathcal{K}$ function. When the parameters are poorly chosen, the system can be conservative, as seen by the large value for $\min h(\bm x)$ in the first row of Table \ref{tab:cmp}
 for a HOCBF with cubic class $\mathcal{K}$ functions. The number of parameters in a HOCBF is at least 5, including $\Delta t, p_1, q_1, p_2, q_2$. If we define polynomial class $\mathcal{K}$ functions, then a HOCBF has an infinite number of parameters. In contrast, the rTLC requires the single parameter $\Delta t$ which has the simple interpretation of directly controlling the rTLC controller conservativeness level.
 }

Finally, the HOCBF and TLC methods do not guarantee forward invariance of the safe set due to the inter-sampling effect while our proposed rTLC does. We need to employ additional methods (e.g., event-triggered methods) to strictly ensure the forward invariance of the HOCBF and TLC methods.

{
To summarize, the two advantages of rTLC are: (i) The fact that it is not subject to inter-sampling effects due to time discretization for implementation purposes, and (ii) Its ease of implementation, since it does not involve any class $\mathcal{K}$ function and requires only a single tunable parameter $\Delta t$ through which the level of conservativeness is also adjusted.
}

 

\begin{table}
\vspace{-2mm}
\caption{Comparison of minimum CBF values and computation time between time-driven HOCBF, time-driven TLC, event-driven TLC, and the proposed rTLC. Safety in time-driven TLC is violated due to the inter-sampling effect. rTLC-1: $(\Delta t,dt)= (0.85,0.1)$, rTLC-2: $(\Delta t,dt) = (0.5,0.5)$, rTLC-3: $(\Delta t,dt) = (0.1,0.1)$}
\label{tab:cmp}
\begin{tabular}{p{2.8cm}<{\centering}p{1.0cm}<{\centering}p{2.2cm}<{\centering}p{0.9cm}<{\centering}r}
\toprule
Method & min $h(\bm x)$  & Compute time (s) & num. of param. \\
\midrule
HOCBF (poor-tuned) 
& 0.5513 & 0.0033$\pm$1.6224e-4 & $\geq 5$ \\
HOCBF (fine-tuned) 
& 1.825e-7 & 0.0035$\pm$2.1814e-4 & $\geq 5$ \\
Time-driven TLC 
     & -0.6414  & 0.0033$\pm$7.5560e-5 & 1\\
     
Event-driven TLC
& 0.6511 & 0.0043$\pm$1.5088e-4 &  3\\ 
rTLC-1
& 6.2487 & 0.0043$\pm$3.3975e-4 & 1\\
rTLC-2
& 1.1865 & 0.0033$\pm$1.4121e-4 & 1\\
rTLC-3
& 0.0125  & 0.0033$\pm$1.0281e-4 & 1
\\\bottomrule
\end{tabular}
\end{table}

\section{Conclusion}
\label{sec:conclusion}

We proposed a robust Taylor-Lagrange control method for ensuring the safety of nonlinear control systems using a higher-order expansion of the safety function with Taylor's theorem with Lagrange remainder. The proposed robust Taylor-Lagrange control method can naturally address the inter-sampling effect 
{and guaranteeing forward invariance while requiring a single hyper-parameter compared to several more needed for other methods.
}We applied the proposed method to an adaptive cruise control problem to demonstrate the effectiveness of our methods. 
In the future, we plan to study the tradeoff between conservativeness and feasibility (in terms of the inter-sampling effect and the optimization feasibility under control bounds) of the proposed robust Taylor-Lagrange control method. 
We will also investigate how to compute a tighter reachable set for the Lagrange remainder to reduce the conservativeness of the proposed robust Taylor-Lagrange control method, as well as employ receding horizon optimization methods (e.g., MPC) for the rTLC method to ensure optimality.

\bibliographystyle{IEEEtran}
\bibliography{HOCBF}

@book{taylor1717methodus,
  title={Methodus incrementorum directa \& inversa},
  author={Taylor, Brook},
  year={1717},
  publisher={Inny}
}

@article{xiao2025taylor,
  title={Taylor-Lagrange Control for Safety-Critical Systems},
  author={Xiao, Wei and Li, Anni},
  journal={arXiv preprint arXiv:2512.11999},
  year={2025}
}

@book{Lagrange1797,
  title={Théorie des functions analytiques},
  author={Joseph-Louis Lagrange},
  year={1797},
  publisher={De l'Imprimerie de la Republique}
}

@article{Ames2017,
	author        = "A. D. Ames and X. Xu and J. W. Grizzle and P. Tabuada",
	title         = "Control Barrier Function Based Quadratic Programs for Safety Critical Systems",
	journal     = "IEEE Transactions on Automatic Control",
	volume        = "62",
	number        = "8",
	year          = "2017",
	pages         = "3861-3876"
}

@article{althoff2014online,
  title={Online verification of automated road vehicles using reachability analysis},
  author={Althoff, Matthias and Dolan, John M},
  journal={IEEE Transactions on Robotics},
  volume={30},
  number={4},
  pages={903--918},
  year={2014},
  publisher={IEEE}
}

@inproceedings{asarin2003reachability,
  title={Reachability analysis of nonlinear systems using conservative approximation},
  author={Asarin, Eugene and Dang, Thao and Girard, Antoine},
  booktitle={International Workshop on Hybrid Systems: Computation and Control},
  pages={20--35},
  year={2003},
  organization={Springer}
}

@inproceedings{cheng2019end,
  title={End-to-end safe reinforcement learning through barrier functions for safety-critical continuous control tasks},
  author={Cheng, Richard and Orosz, G{\'a}bor and Murray, Richard M and Burdick, Joel W},
  booktitle={Proceedings of the AAAI conference on artificial intelligence},
  volume={33},
  number={01},
  pages={3387--3395},
  year={2019}
}

@inproceedings{Nagumo1942berDL,
	title={{\"U}ber die Lage der Integralkurven gew{\"o}hnlicher Differentialgleichungen},
	author={Mitio Nagumo},
	booktitle={Proceedings of the Physico-Mathematical Society of Japan. 3rd Series. 24:551-559},
	volume={},
	number={},
	year={1942}
}

@article{clark2021control,
  title={Control barrier functions for stochastic systems},
  author={Clark, Andrew},
  journal={Automatica},
  volume={130},
  pages={109688},
  year={2021},
  publisher={Elsevier}
}

@inproceedings{srinivasan2018control,
  title={Control of multi-agent systems with finite time control barrier certificates and temporal logic},
  author={Srinivasan, Mohit and Coogan, Samuel and Egerstedt, Magnus},
  booktitle={2018 IEEE Conference on Decision and Control (CDC)},
  pages={1991--1996},
  year={2018},
  organization={IEEE}
}

@article{xiao2021adaptive,
  title={Adaptive control barrier functions},
  author={Xiao, Wei and Belta, Calin and Cassandras, Christos G},
  journal={IEEE Transactions on Automatic Control},
  volume={67},
  number={5},
  pages={2267--2281},
  year={2021},
  publisher={IEEE}
}

@article{xiao2022event,
  title={Event-triggered control for safety-critical systems with unknown dynamics},
  author={Xiao, Wei and Belta, Calin and Cassandras, Christos G},
  journal={IEEE Transactions on Automatic Control},
  volume={68},
  number={7},
  pages={4143--4158},
  year={2022},
  publisher={IEEE}
}

@article{taylor2020safety,
  title={Safety-critical event triggered control via input-to-state safe barrier functions},
  author={Taylor, Andrew J and Ong, Pio and Cort{\'e}s, Jorge and Ames, Aaron D},
  journal={IEEE Control Systems Letters},
  volume={5},
  number={3},
  pages={749--754},
  year={2020},
  publisher={IEEE}
}

@article{xiao2021high,
  title={High-order control barrier functions},
  author={Xiao, Wei and Belta, Calin},
  journal={IEEE Transactions on Automatic Control},
  volume={67},
  number={7},
  pages={3655--3662},
  year={2021},
  publisher={IEEE}
}

@inproceedings{taylor2020adaptive,
  title={Adaptive safety with control barrier functions},
  author={Taylor, Andrew J and Ames, Aaron D},
  booktitle={2020 American Control Conference (ACC)},
  pages={1399--1405},
  year={2020},
  organization={IEEE}
}

@book{Boyd2004,
	author        = "Stephen P Boyd and Lieven Vandenberghe",
	title         = "Convex optimization",
	publisher     = "Cambridge university press",
	address       = "New York",
	year          = "2004"
}

@article{Tee2009,
	author        = "Keng Peng Tee and Shuzhi Sam Ge and Eng Hock Tay",
	title         = "Barrier lyapunov
	functions for the control of output-constrained nonlinear systems",
	journal       = "Automatica",
	volume        = "45",
	number        = "4",
	year          = "2009",
	pages         = "918-927"
}

@inproceedings{Wieland2007,
	author        = "Peter Wieland and Frank Allgower",
	title         = "Constructive safety using control barrier functions",
	booktitle     = "Proc. of 7th IFAC Symposium on
	Nonlinear Control System",
	year          = "2007",
}

@book{Aubin2009,
	author        = "Jean Pierre Aubin",
	title         = "Viability theory",
	publisher     = "Springer",
	year          = "2009"
}

@article{Prajna2007,
	author        = "Stephen Prajna and Ali Jadbabaie and George J. Pappas",
	title         = "A framework
	for worst-case and stochastic safety verification using barrier certificates",
	journal       = "IEEE Transactions on Automatic Control",
	volume        = "52",
	number        = "8",
	year          = "2007",
	pages         = "1415-1428"
}

@inproceedings{Wisniewski2013,
	author        = "Rafael Wisniewski and Christoffer Sloth",
	title         = "Converse barrier certificate theorem",
	booktitle     = "Proc. of 52nd IEEE Conference on Decision and Control",
	year          = "2013",
	address       = "Florence, Italy",
	pages         = "4713-4718"
}

@article{Glotfelter2017,
	author        = "P. Glotfelter and J. Cortes and M. Egerstedt",
	title         = "Nonsmooth barrier functions with applications to multi-robot systems",
	journal       = "IEEE control systems letters",
	volume        = "1",
	number        = "2",
	year          = "2017",
	pages         = "310-315"
}

@book{Khalil2002,
	author        = "Hassan K. Khalil",
	title         = "Nonlinear Systems",
	edition       = "Third",
	publisher     = "Prentice Hall",
	address       = "",
	year          = "2002"
}

@inproceedings{Nguyen2016,
	author        = "Quan Nguyen and Koushil Sreenath",
	title         = "Exponential Control Barrier Functions for Enforcing High Relative-Degree Safety-Critical Constraints",
	booktitle     = "Proc. of the American Control Conference",
	year          = "2016",
	pages         = "322-328"
}

@book{Bryson1969,
	author        = "Bryson and Ho",
	title         = "Applied Optimal Control",
	publisher     = "Ginn Blaisdell",
	address       = "Waltham, MA",
	year          = "1969"
}

@book{kirk2004optimal,
  title={Optimal control theory: an introduction},
  author={Kirk, Donald E},
  year={2004},
  publisher={Courier Corporation}
}

@article{bellman1966dynamic,
  title={Dynamic programming},
  author={Bellman, Richard},
  journal={Science},
  volume={153},
  number={3731},
  pages={34--37},
  year={1966},
  publisher={American Association for the Advancement of Science}
}

@book{Denardo2003,
	author        = "E. V. Denardo",
	title         = "Dynamic Programming: Models and Applications",
	publisher     = "Dover Publications",
	year          = "2003"
}

@article{garcia1989model,
  title={Model predictive control: Theory and practice—A survey},
  author={Garcia, Carlos E and Prett, David M and Morari, Manfred},
  journal={Automatica},
  volume={25},
  number={3},
  pages={335--348},
  year={1989},
  publisher={Elsevier}
}

@book{Rawlings2018,
	author        = "James B. Rawlings and David Q. Mayne and Moritz M. Diehl",
	title         = "Model Predictive Control: Theory, Computation, and Design",
	publisher     = "Nob Hill Publishing",
	address       = "",
	year          = "2018"
}

\end{document}